%% file: atha.tex
\documentstyle[11pt,iau208,twoside]{article}
\input iau208_epsf.tex

\def\edcomment#1{\iffalse\marginpar{\raggedright\sl#1\/}\else\relax\fi}
\reversemarginpar

\begin{document}
\title{Isolated and Interacting Galaxies : Simulations with GRAPE}
\author{E. Athanassoula}
\affil{Observatoire de Marseille, 2 place Le Verrier, 13248 Marseille
 cedex 04, France}

\begin{abstract}
I present $N$-body simulations of isolated and interacting
galaxies, made on GRAPE machines. In particular I discuss the formation
and evolution of 
$N$-body bars and compare their properties to those of bars in
early-type and late-type galactic discs. I argue that the halo
can help the bar grow, contrary to previous beliefs, by taking 
positive angular momentum from it via its resonant stars. I 
then focus on the
interaction and subsequent merging of a barred disc galaxy with a
spheroidal satellite. The evolution depends strongly on the mass 
(density) of the satellite and may lead to its destruction or to the
destruction of the bar.
\end{abstract}

\section{Introduction}

GRAPE machines (Makino, these proceedings) have given a substantial
boost to the study of galactic dynamics by making state of the art
$N$-body simulations possible at a relatively low cost and by allowing a
simulation 
environment which is much more flexible than that of supercomputing
centers. Machines with even numbers, e.g. GRAPE-4 (Makino et al. 1997)
and GRAPE-6 (Makino, these proceedings), have a 
very high accuracy arithmetic and calculate also the derivative of the
force, thus allowing a more accurate integration scheme to follow the
evolution, as 
well as calculations with little or no softening. In the field of
galactic dynamics they are particularly
suited for studying the nuclei of galaxies, the effect of black holes,
as well as for problems where softening should be limited or avoided, or for
other cases where high precision is necessary. Machines with odd
numbers, as GRAPE-3 (Ebisuzaki et al. 1993) and GRAPE-5 (Kawai
et al. 2000), have a more limited accuracy and do
not provide the derivatives of the force. Nevertheless the accuracy of
GRAPE-3 was shown to be amply sufficient for a large number of problems on
galactic dynamics (e.g. Athanassoula et al. 1998) and GRAPE-5  is even
more accurate (Kawai et al. 2000).
They are thus very well suited for studies of the evolution of
isolated and interacting galaxies, as well as of galaxy systems.
They have therefore been widely
used in these topics, as witnessed by the ever-increasing number of
publications based on GRAPE simulations. The field of research thus
covered is too 
broad to be usefully covered by a single review, so I will limit
myself here to a few, largely unpublished, results obtained
recently with the Marseille GRAPE systems. In
sections~2 and 3 I will present results
on the formation of bars and on the secular evolution of barred disc
galaxies. In section~4 I will discuss the interaction
and subsequent merging of a barred disc galaxy and a small
elliptical companion. A few concluding remarks are given in 
section~5.

\section{Formation and properties of bars in $N$-body discs}
\label{sec:barevol}

\begin{figure}
\plotone{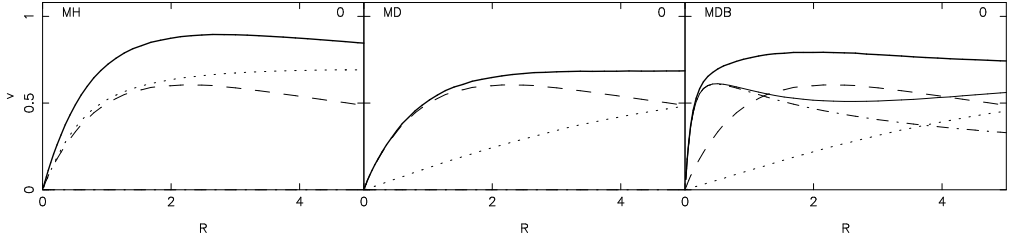}
\caption{Circular velocity curves corresponding to the initial
conditions of the three simulations discussed in section 2 (solid
lines). The contribution of 
the disc, halo and bulge components are given by dashed, dotted and
dot-dashed lines, respectively. The thin solid line in the right panel
gives the total contribution of the two spherical components, halo and
bulge. Here and in subsequent figures the names of the simulations 
are given in the upper left corner
of each panel and the time in the upper right.}
\label{fig:morpho_rotcur}
\end{figure}

\begin{figure}
\plotone{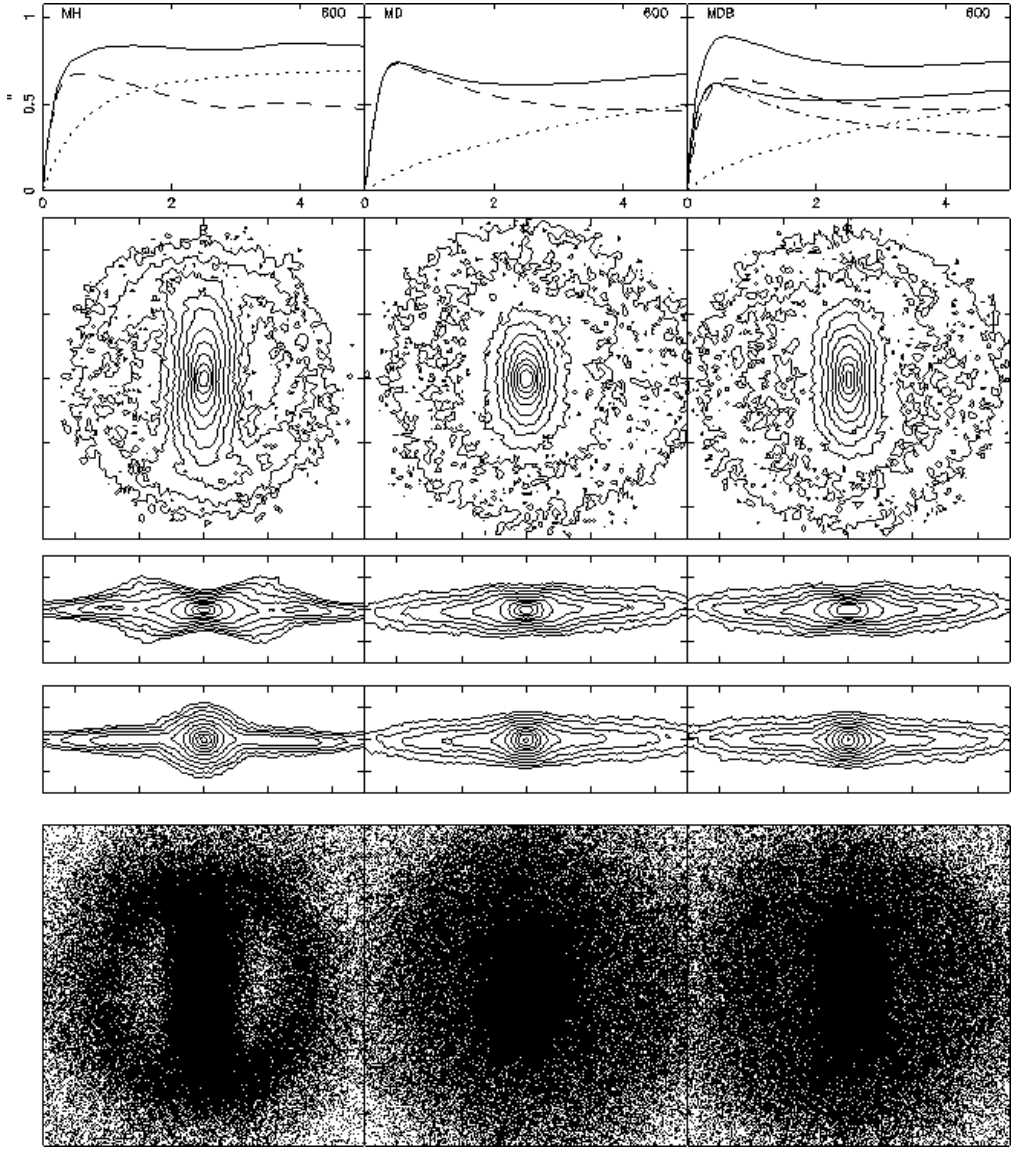}
\caption{Our three fiducial models at time 600. The three upper panels
show the circular velocity curves, and their layout is the same as for
Fig.~\ref{fig:morpho_rotcur}. The next three rows of panels show
isodensity contours of the disc component, seen face-on (second row),
and edge-on with the bar seen side-on (third row) and end-on (fourth
row). The fifth row of panels shows the dot-plots of the face-on 
distributions. The size of the square box 
for the second and fifth row of panels is 10 initial disc scale 
lengths, and all panels, except for the upper row, have the 
same linear scale.
}
\label{fig:basic600}
\end{figure}

Since the early seventies it is well known that galactic discs can be bar
unstable (e.g. Miller, Prendergast \& Quirk 1970; Hohl 1971). 
Nevertheless, the bar instability, and in particular its origin and its
effect on galaxy evolution, are still not well understood (e.g. Sellwood
2000). Let me add here some new elements to this discussion with the help
of three fully self-consistent simulations, run on the Marseille
GRAPE-5 systems. In this section I will introduce the simulations and
discuss the morphology of the ensuing $N$-body bar. In the next
section I will touch on the origin of the instability and in
particular on the role of the halo.

In all three simulations described here the disc component is
an exponential disc of unit scale length and unit mass. The halo
component is initially thermalised, spherical and non-rotating, 
and has a total
mass five times that of the disc. The first two simulations have
haloes of very different central concentrations. This can be seen in
Fig.~\ref{fig:morpho_rotcur}, which shows the initial circular velocity
curves, together with the contribution of each component
separately. We will call the simulation with the concentrated halo MH, for
massive halo, since the halo is somewhat more massive than the disc in
the inner parts, and the second one MD, for massive disc, since now it is
the disc component that dominates in the inner parts. The initial
conditions of the third simulation are 
identical to those of the second one, except that it has also a
Plummer sphere bulge of mass 0.6 and radius 0.4. We will hereafter refer
to it as MDB, for massive disc with bulge. The number of particles is
of the order of 1.2 $\times$ $10^6$. These three initial
conditions are very similar to those described in Athanassoula \&
Misiriotis (2002, hereafter AM), except that the disc here is
initially somewhat thinner. A reasonable calibration (AM) 
gives that $t$ = 500 corresponds to 7 $\times 10^9$ years.  
Different values, however, can be obtained for different scalings of 
the disc mass and scale length.

We let these simulations evolve up to time 900. In all three cases a
bar forms, but its properties are different, as witnessed in
Fig.~\ref{fig:basic600}, which shows results for time 600. The upper
panels show the circular velocity
curves. Comparing them with the initial ones, we see that
in all three cases the disc component is considerably more centrally
concentrated than at time $t$ = 0. In model MD the difference is only
quantitative, since 
the disc dominates in the inner parts both at $t$ = 0 and at 
$t$ = 600, while
for model MH the difference is qualitative, since at $t$ = 600 the
disc dominates in the inner parts, contrary to what was the case
initially. 

We note very large morphological differences between
models MH and MD, model MDB being intermediate. Thus the bar in model
MH is stronger, longer and more rectangular-like than in
model MD. It also has a massive ring surrounding the bar, whose
diameter is roughly equal to the bar major axis, as is the case for
inner rings in real galaxies (Kormendy 1979, Buta 1986). It is
somewhat elongated 
along the bar, in good agreement with observations of inner rings 
(Buta 1995). Model MDB has a bar of intermediate length and shape.

Let us now compare the properties of $N$-body bars with those of bars
in early and 
late type galaxies. Bars in early type galaxies are longer than those
in late type ones (Elmegreen \& Elmegreen 1985) and have
rectangular-like isophotes (Athanassoula et al. 1990), as the bar in
model MH. Furthermore, the
bar in MH-type simulations often has ansae, a feature
seen only in early type barred galaxies (good examples are shown on 
page 42 of the Hubble Atlas of Galaxies, Sandage 1961). Thus the bar in
MH-type models can be compared to a bar in an early type galaxy, while 
that in MD-type models with a late type. 

The edge-on views of the three models are also quite different. Model
MD shows a boxy 
shape when viewed side-on, while model MH has a strong peanut. Such
shapes are indeed observed in many edge-on galaxies and are
linked to the existence of a bar in the disc (Kuijken \&
Merrifield 1995, Bureau \& Athanassoula 1999, Athanassoula \& Bureau
1999).

\begin{figure}
\plotone{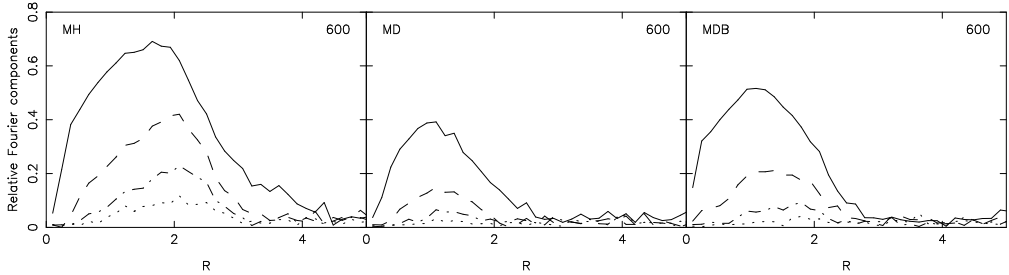}
\caption{Relative Fourier components of the face-on density
distribution. The $m$ = 2, 4, 6 and 8 components are plotted with a
solid, dashed, dot-dashed and dotted line respectively. }
\label{fig:mcompon600}
\end{figure}

Fig.~\ref{fig:mcompon600} shows the relative $m$ = 2, 4, 6 and 8 Fourier
components of 
the face-on density, for the three models. We note that they 
are much larger for model MH, as could be expected since the bar is
stronger. Let me stress that the $m$ = 6 and 8 components
are in the noise for model MD, while they are important for model
MH. Note also that the $m$ = 4 and 6 components in model MH have 
relative amplitudes of the same order as the $m$ = 2 in model MD. Such Fourier
components have also been calculated for real barred galaxies. In his
review, Ohta (1996) describes them both for early and for late type
galaxies and his descriptions fit well models MH and MD
respectively. Thus he stresses that for early type galaxies the
relative even Fourier components are large and the higher $m$ components
are important. On the contrary for late type
bars all relative Fourier components have lower amplitude, while the high
$m$ components are negligible (cf. his Fig.~2). 

\begin{figure}
\plotone{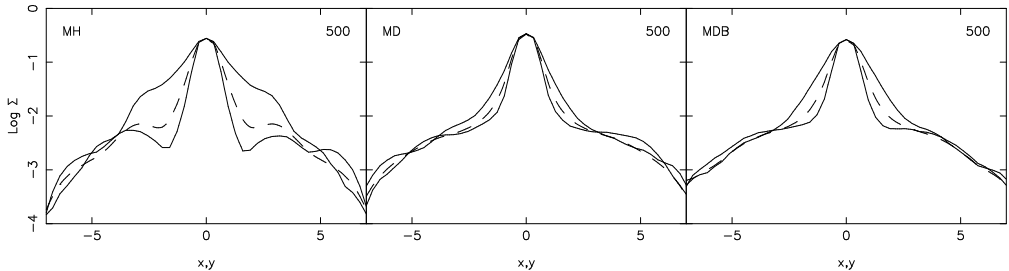}
\caption{Projected surface density of the disc component viewed
face-on. The solid lines correspond to cuts along the major and
minor axes of the bar. The dashed line is obtained by azimuthal averaging.}
\label{fig:faceprof}
\end{figure}

Let us now try and compare the results of the $N$-body simulations
with photometric results. For this we view the galaxy face-on and 
obtain the projected density profile along the directions of the bar
major and minor axes (Fig.~\ref{fig:faceprof}). We note that
for model MH the profile along the bar major axis has flat parts on
either side of the center, followed by steep drops at the ends of the
bar. The picture is totally different for model MD, where the profile
drops clearly all through the bar region, and there is no clear
change of slope at the end of the bar. Observers have made similar cuts
for many barred galaxies. Elmegreen \& Elmegreen (1985) classify these
profiles in two types : {\it flat}, which are reminiscent of the profile of
model MH, and {\it exponential}, which are similar to that of model MD. 
They stress that flat profiles are found in early type galaxies,
while exponential ones mainly in late types. Similar conclusions were
reached by Ohta, Hamabe \& Wakamatsu (1990), Ohta (1996), etc.
Thus again model MH resembles bars in early type galaxies and model
MD late type ones. 

\begin{figure}
\plotone{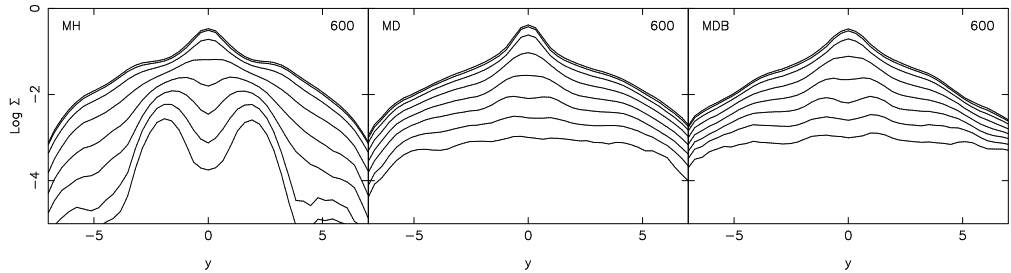}
\caption{Projected surface density of the disc component viewed
edge-on, with the bar seen side-on. The various profiles are obtained
from cuts at various distances from the equatorial plane. The top-most
one is for $z$ = 0 and the rest are spaced by $\Delta z$ = 0.2
apart. Thus the lower-most one corresponds to $z$ = 1.4.}
\label{fig:edgeonprof600}
\end{figure}

Let us repeat the exercise, now viewing the galaxy edge-on and the bar
side-on. The results are shown in Fig.~\ref{fig:edgeonprof600}. For
model MH the
cuts at high $z$ values show a minimum at the center, followed by two
maxima, one on either side, and then a steep drop. This is the clear
signature of a peanut. Similar features, but of {\it much} lower amplitude,
can also be seen for model MD, revealing the existence of a weak
peanut or a boxy feature. For model MH, the cuts at $z$ = 0 show a
ledge on either 
side of the nucleus, followed by a relatively sharp drop. L\"utticke et
al. (2000) have made similar profiles for a large number of edge-on
galaxies and their profiles are in good agreement with those of
Fig.~\ref{fig:edgeonprof600}. They also find a correlation, in the
sense that it 
is the strongest peanuts that have the most pronounced ledges,
i.e. the strongest bars. This could be predicted from the simulations,
both those described here, and others with somewhat different sets of
initial parameters. If models of type MH are indeed associated with
early type galaxies, then the simulations predict that the strongest 
peanuts will form in early types. This is unfortunately difficult to
verify since it is difficult to find the morphological type of an edge-on disc
galaxy.

\begin{figure}
\plotone{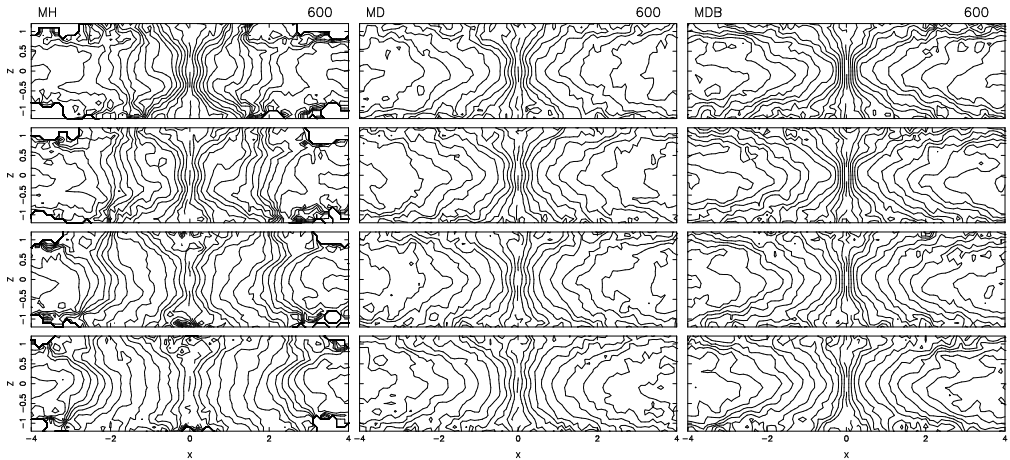}
\caption{Velocity field of the disc component seen edge-on. In the
upper panels the bar is seen end-on and in the lower ones side-on. The
second and third row of panels are for intermediate viewing angles, at
30 and 60 degrees from the bar major axis, respectively.}
\label{fig:edgeonv2600}
\end{figure}

Fig.~\ref{fig:edgeonv2600} shows the velocity field of the models seen
edge-on. We note that when the viewing angle is along the bar minor
axis, i.e. when the bar is viewed side-on, the velocities within 
the bar/peanut region do not change much with distance from the 
equatorial plane. This behaviour is often termed `cylindrical
rotation' and was observed for many peanut galaxies (e.g. NGC 4565, 
Kormendy \& Illingworth 1982; NGC 3079, Shaw, Wilkinson \& Carter
1993; NGC 128, D'Onofrio et al. 1999). 

In all the above comparisons model MDB is intermediate between models
MH and MD. This holds for the length of the bar and the shape of its
face-on isophotes, as well as for its Fourier components and the
strength of its peanut.  
For this model we also analysed the form of the bulge. This starts
initially as spherical but evolves to
oblate, with its shortest dimension perpendicular to the
equatorial plane. The departures from sphericity are strongest for the
innermost isophotes, which are also triaxial. Here we find axial
ratios of the order of 0.7 and 0.75. At intermediate radii we find
ratios of the order of 0.9 and 0.8, while in the outer bulge parts the
form of the isophotes approaches a sphere.

It is also worth adding a few words about the inner parts of the
bar/disc component in model MH. If the bar is seen edge-on and the bar
end-on (Fig.~\ref{fig:basic600}, panel in fourth row and left column),
then the bar presents a structure which could be easily
mistaken for an intermediate size bulge in an edge-on disc
galaxy. This impression is reinforced if we look at the projected
density profiles of the $N$-body system. Such `bulges' could be
erroneously included in observational studies of bulges in edge-on
disc systems. 

Summarising this section I note that the results of the $N$-body 
simulations reproduce 
well the morphology, the photometry and the kinematics of barred
galaxies. They also produce strong arguments linking early type barred
galaxies with simulations in which the halo is {\it initially}
centrally concentrated and late type galaxies with simulations where
the center part is {\it initially} disc dominated. 

\section{The role of the halo}
\label{sec:halo}

\begin{figure}
\plotfiddle{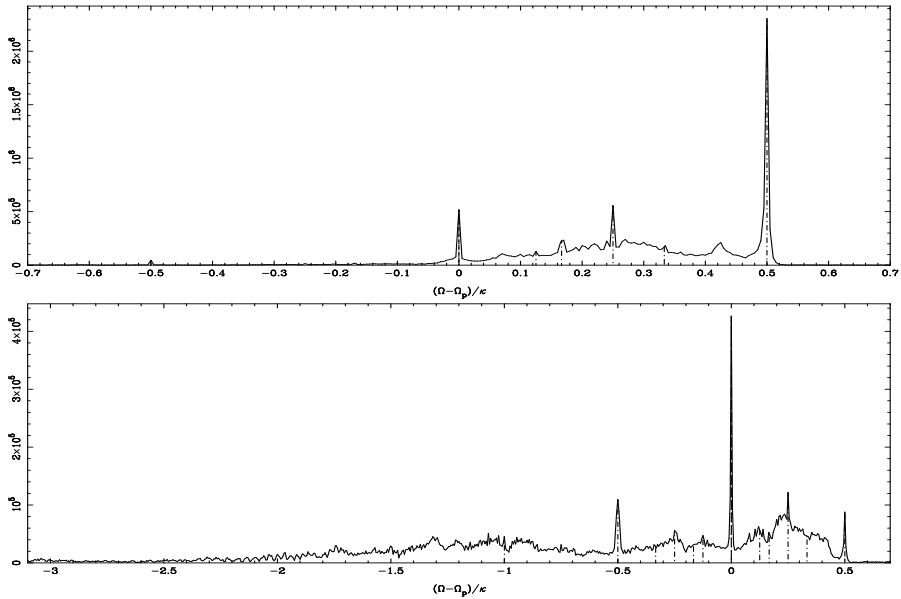}{7.9cm}{270}{50}{40}{-200}{250}
\caption{Number density of particles as a function of the frequency
ratio $(\Omega - \Omega_p) / \kappa$ 
for model MH and time 500. The upper panel 
corresponds to the disc particles 
and the lower one to the halo ones. The dot-dashed vertical lines mark
the positions of the main resonances.}
\label{fig:spectrum}
\end{figure}

In the above we saw that the strongest bar formed in the case where the
halo contribution in the disc region is initially the strongest. 
In order to understand this I froze
the potential of both models MH and MD at time $t$ = 500 and calculated orbits
of 200\,000 particles chosen at random from the simulation particles
-- half taken from the disc and half from the halo --. For each orbit
I calculated the basic frequencies $\Omega$, $\kappa$ and $\kappa_z$,
i.e. the angular, epicyclic and vertical frequencies. This is not always
a straightforward calculation. Although the results for of the order
of 10\% of the particles are not very reliable, in particular for 
$\Omega$, the remainder are precise enough to warrant discussion.

Fig.~\ref{fig:spectrum} shows the number density of particles/orbits 
that have a given 
value of ($\Omega - \Omega_p) / \kappa$ as a function of this ratio, 
for run MH. We
note that the distribution is far from homogeneous and that there are
large peaks at the main resonances. This could have been expected for 
the disc particles, although the sharpness and the amplitude of the 
resonance peaks is worth noting. What is more novel and exciting, 
however, is the existence and 
the amplitude of the resonant peaks in the halo population, which 
has often been considered as a non- or little-responsive component. 
For the disc particles the highest 
peak is at the Inner Lindblad resonance (ILR), followed by corotation
(CR) and inner Ultra-harmonic or 1:4 resonances. For the halo the 
highest peak is at CR, followed by the outer 
Lindblad resonance (OLR) and then the ILR. The amplitude of the 
peaks varies with 
time, but also from one simulation to another, as will be discussed 
in detail elsewhere. 

Disc stars at resonances can give or take angular momentum, and 
thus contribute to the evolution of the system. Stars at ILR emit 
angular momentum, while stars at CR or OLR absorb it (Lynden-Bell 
\& Kalnajs 1972). The resonant halo stars 
generally absorb angular momentum (Tremaine \& Weinberg 1984). 
Thus, together with 
the disc stars at CR and OLR, they will destabilise the bar and induce 
it to increase its amplitude, since the bar has negative energy and 
angular momentum. This can also be seen in a simpler way 
if we calculate the angular momentum of the two components (not 
plotted here). A more explicit and quantitative treatment of this 
exchange will be given elsewhere, but the short discussion above 
can give some insight as to why a halo may help the bar grow stronger, 
instead of stabilising it.
 
\section{Merging of a barred galaxy with a satellite}
\label{sec:interact}

GRAPE is particularly useful for simulations of interacting systems,
since its force calculation does not depend on geometry, contrary to
grid or expansion methods. The simulations I will present here
describe the interaction and subsequent merging of a barred disc
galaxy with its spherical satellite. The target galaxy is composed of
a disc and a halo component with a mass ratio $M_h/M_d$ = 0.54. The
satellite is modeled with a Plummer sphere of mass $M_s$
and is initially placed near the edge of the halo in a
direct near-circular orbit in the disc plane. We consider two
different companion masses, one equal to the mass of the disc and the
other one tenth of that. Since the scale lengths are the same in the two
cases, the ratio of masses reflects also the ratio of densities. 

Due to dynamical friction, the companion spirals inwards and reaches
the center of the target galaxy. The rate at which it does so depends
heavily on its mass. Thus the high mass companion
reaches the center in one third of the time taken by the low mass
companion. As a result of this evolution the target disc thickens, but
also expands, so that its aspect ratio, when seen edge-on, does not
change much, staying that of a disc (cf. Fig. 4 of Athanassoula
1996). The companion, occupying the central region, forms a bulge, or
contributes to one (cf. also Walker, Mihos \& Hernquist 1996). Its shape 
is oblate and it has lost only a small fraction of its initial mass.

We thus have an evolution of the target galaxy from late to early
type. If the plane of the initial companion orbit is at an angle to
that of the disc, then the target disc still thickens and expands, but
it also tilts, so that the final equatorial plane is, for the case of
the massive companion, not far from that of the companion's initial
orbital plane. For the low mass companion it tilts much less, but
still the tilt is noticeable. The energy of the vertical motion of the
companion, due to the tilt,  is transformed mainly to an organised
motion of the disc particles, rather than to an increase of the
vertical component of the velocity dispersion, so that the disc is not
excessively thickened (cf. also Athanassoula 1996, Huang \& Carlberg
1997).

\begin{figure}
\plotone{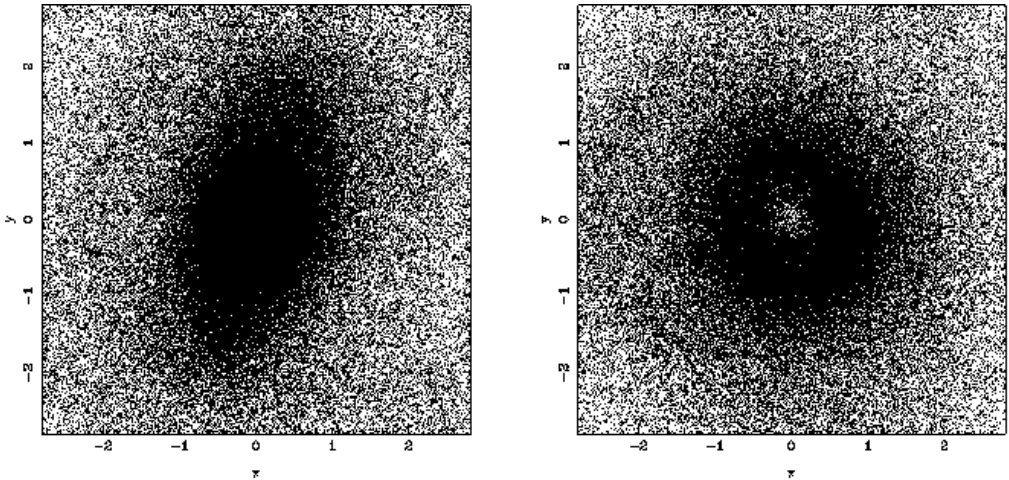}
\caption{Face-on view of the disc particles at the beginning of the
  simulation (left panel) and after the merging with the high mass
  companion (right panel). As a
  result of the merging the initial bar has disappeared and a low
  density region has formed in the central parts, occupied by the companion
  particles (not plotted here).}
\label{fig:discdens}
\end{figure}

The high mass companion exerts a strong perturbation on the bar, which
of course becomes stronger as the companion approaches the central
regions of the target galaxy. Particles are drawn from the bar in the
direction of the companion, so that the bar is progressively emptied,
loses its identity and by the time the companion reaches the center
the bar has wrapped around it. Thus after the merging the target is a
non-barred galaxy and its disc projected surface density has a minimum
in the central 
region (left panel of Fig.~\ref{fig:discdens}), which is now
occupied by the companion (cf. Fig. 2 of Athanassoula 1996). 

\begin{figure}
\plotone{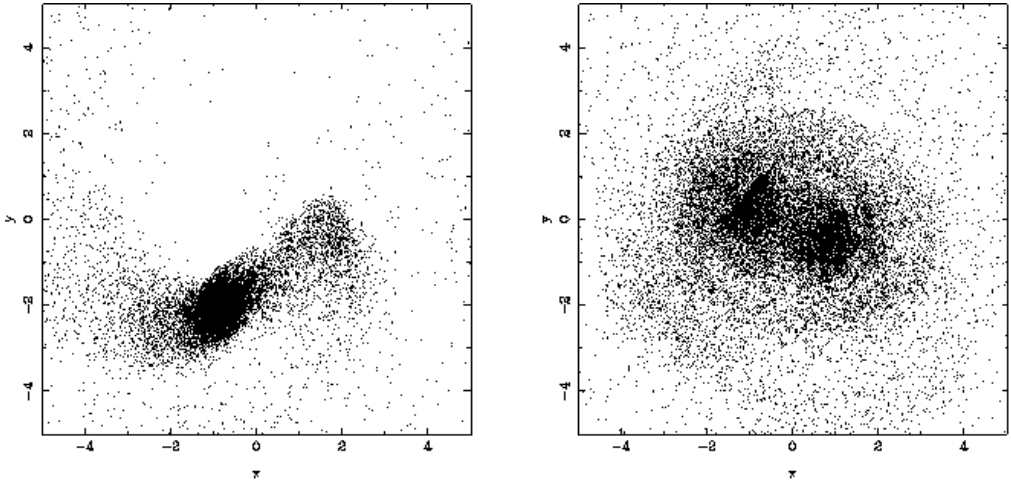}
\caption{Fate of the low mass companion during the 
  merging. The companion loses particles
  preferentially from the direction if its motion and the
  diametrically opposite direction, and becomes elongated along the
  orbit (left panel). The companion is thus
  disrupted and, after the merging is completed, its particles orbit
  around the bar of the target galaxy, 
  extending and elongating it (right panel).}
\label{fig:companion}
\end{figure}

The evolution is quite different in the case with the low mass
companion. Now it is the companion that gets most perturbed,
particularly after it has reached the disc region. It loses particles, not
uniformly from all its surface, but from two preferred directions, one
in the direction of the motion and the other opposite to it. As it
does so its mass decreases and it also 
loses its initial spherical shape and becomes strongly elongated in
the direction of the orbital motion
(left panel of Fig.~\ref{fig:companion}). When it reaches the bar region, the
companion loses particles at a yet higher rate, until it
disintegrates. Its particles then orbit around the bar in elongated
orbits, staying longer near the bar major axis. At any moment
there is a concentration of companion particles around the two
opposite ends of the bar, thus extending and lengthening it
(right panel of Fig.~\ref{fig:companion}). These concentrations are
not transient.  

The in-falling satellite perturbs also the halo component. The most
massive companion induces a considerable spin in the direction of
the initial 
rotation of the companion. Furthermore the halo velocity dispersion
increases locally at the location of the  companion. This creates 
a local maximum of the velocity dispersion, which moves inwards 
with the satellite. The velocity
dispersion after the companion has reached the center is larger than
the initial one, particularly in the central parts. The heating of the
disc has been examined by Walker et al. (1996), Huang \& Carlberg 
(1997) and Vel\`azquez \& White (1999).

\section{Summary}
\label{sec:conclusions}

In the above I presented results of simulations made with the
Marseille GRAPE systems. I started with simulations of the formation
and evolution of a bar in an isolated disc galaxy. These show that
strong bars can form in systems with a sizeable halo component within
the disc region. I find very good agreement between the properties of
the $N$-body bars and those of bars in galaxies. In particular I
presented arguments that link $N$-body bars that grew in a
environment with a considerable halo component with early type bars,
and bars growing in a disc-dominated environment with late type
bars. Such simulations stress the role of the halo in the development
of the bar instability and show that the halo can stimulate bar
formation, contrary to the common belief that it will always quench
it. The reason is that the halo responds to the bar and a considerable
fraction of its particles are in resonance with it. This of course
would have been missed by simulations treating the halo as a rigid
component. 

I then discussed the interaction and subsequent merging of a barred
disc galaxy with its spherical satellite. If the satellite has high
mass (density), then it will spiral rapidly towards the center of the
disc, destroy the bar and occupy the center of the target, thus
forming a bulge, or contributing to one. Such interactions will cause
evolution along the Hubble sequence, from late to early type
discs. During this process the disc thickens considerably, but also
expands, so that its final shape is still that of a disc, albeit
somewhat thicker than the initial one.
If the companion has low mass (density), then it will get disrupted as
it spirals inwards towards the center of the target.
If the orbit of the companion is initially at an angle with the plane
of the disc, then the disc tilts but is not destroyed. In the case of
the high mass (density) companion, the tilt angle is not far from that
of the plane of the initial orbit of the companion.

The above results, and many others not shown here, argue that GRAPE is
an ideal tool for simulations of galaxies and galaxy systems.
In all the above simulations
the number of particles was of the order of a
million, or more than that. This is now a standard number for most of
our GRAPE simulations. GRAPE can of course handle considerably larger
numbers, provided its front end has a sufficient memory, and
we have used such numbers whenever necessary.  

GRAPE boards permit fast and accurate calculations of the gravitational 
potential and forces, and are not limited by the specific geometry 
of the particle distribution.
Their price is a very small fraction of that of
supercomputers. Both high precision direct summation codes and tree
codes have been developed for them. They offer a flexible simulation 
environment. They are
thus ideally suited for small research groups ready to 
invest themselves in the GRAPE approach.  
The close collaboration between the members of the
international community of GRAPE users contributes further to the
success of these machines.

\acknowledgments

First and foremost I would like to thank the Tokyo group -- headed
initially by D. Sugimoto and later by J. Makino -- for developing the
GRAPE systems, without which the simulations presented here
would not have been possible.
I would like to thank Albert Bosma for many discussions on barred
galaxies and A. Misiriotis for his collaboration on the software
calculating the orbital frequencies. I
would also like to thank the IGRAP, the Region PACA, the
INSU/CNRS and the University of Aix-Marseille I for funds to develop
our GRAPE computing facilities.

\end{document}

%% file: iau208_epsf.tex
\ifx\epsfannounce\undefined \def\epsfannounce{\immediate\write16}\fi
 \epsfannounce{This is `epsf.tex' v2.7k <10 July 1997>}%
\newread\epsffilein    
\newif\ifepsfatend     
\newif\ifepsfbbfound   
\newif\ifepsfdraft     
\newif\ifepsffileok    
\newif\ifepsfframe     
\newif\ifepsfshow      
\epsfshowtrue          
\newif\ifepsfshowfilename 
\newif\ifepsfverbose   
\newdimen\epsfframemargin 
\newdimen\epsfframethickness 
\newdimen\epsfrsize    
\newdimen\epsftmp      
\newdimen\epsftsize    
\newdimen\epsfxsize    
\newdimen\epsfysize    
\newdimen\pspoints     
\def\epsfbox#1{\global\def\epsfllx{72}\global\def\epsflly{72}%
   \global\def\epsfurx{540}\global\def\epsfury{720}%
   \def\lbracket{[}\def\testit{#1}\ifx\testit\lbracket
   \let\next=\epsfgetlitbb\else\let\next=\epsfnormal\fi\next{#1}}%
\def\epsfgetlitbb#1#2 #3 #4 #5]#6{%
   \epsfgrab #2 #3 #4 #5 .\\%
   \epsfsetsize
   \epsfstatus{#6}%
   \epsfsetgraph{#6}%
}%
\def\epsfnormal#1{%
    \epsfgetbb{#1}%
    \epsfsetgraph{#1}%
}%
\newhelp\epsfnoopenhelp{The PostScript image file must be findable by
TeX, i.e., somewhere in the TEXINPUTS (or equivalent) path.}%
\def\epsfgetbb#1{%
    \openin\epsffilein=#1
    \ifeof\epsffilein
        \errhelp = \epsfnoopenhelp
        \errmessage{Could not open file #1, ignoring it}%
    \else                       
        {
            \chardef\other=12
            \def\do##1{\catcode`##1=\other}%
            \dospecials
            \catcode`\ =10
            \epsffileoktrue         
            \epsfatendfalse     
            \loop               
                \read\epsffilein to \epsffileline
                \ifeof\epsffilein 
                \epsffileokfalse 
            \else                
                \expandafter\epsfaux\epsffileline:. \\%
            \fi
            \ifepsffileok
            \repeat
            \ifepsfbbfound
            \else
                \ifepsfverbose
                    \immediate\write16{No BoundingBox comment found in %
                                    file #1; using defaults}%
                \fi
            \fi
        }
        \closein\epsffilein
    \fi                         
    \epsfsetsize                
    \epsfstatus{#1}%
}%
\def\epsfclipoff{\def\epsfclipstring{\ifepsfdraft\space clip\fi}}%
\epsfclipoff 
\def\epsfspecial#1{%
     \epsftmp=10\epsfxsize
     \divide\epsftmp\pspoints
     \ifnum\epsfrsize=0\relax
       \includegraphics{\ifepsfdraft}%
     \else
       \epsfrsize=10\epsfysize
       \divide\epsfrsize\pspoints
       \includegraphics{\ifepsfdraft}%
     \fi
}%
\def\epsfframe#1%
{%
  \leavevmode                   
  \setbox0 = \hbox{#1}%
  \dimen0 = \wd0                                
  \advance \dimen0 by 2\epsfframemargin         
  \advance \dimen0 by 2\epsfframethickness      
  \vbox
  {%
    \hrule height \epsfframethickness depth 0pt
    \hbox to \dimen0
    {%
      \hss
      \vrule width \epsfframethickness
      \kern \epsfframemargin
      \vbox {\kern \epsfframemargin \box0 \kern \epsfframemargin }%
      \kern \epsfframemargin
      \vrule width \epsfframethickness
      \hss
    }
    \hrule height 0pt depth \epsfframethickness
  }
}%
\def\epsfsetgraph#1%
{%
   \leavevmode
   \hbox{
     \ifepsfframe\expandafter\epsfframe\fi
     {\vbox to\epsfysize
     {%
        \ifepsfshow
            \vfil
            \hbox to \epsfxsize{\epsfspecial{#1}\hfil}%
        \else
            \vfil
            \hbox to\epsfxsize{%
               \hss
               \ifepsfshowfilename
               {%
                  \epsfframemargin=3pt 
                  \epsfframe{{\tt #1}}%
               }%
               \fi
               \hss
            }%
            \vfil
        \fi
     }%
   }}%
   \global\epsfxsize=0pt
   \global\epsfysize=0pt
}%
\def\epsfsetsize
{%
   \epsfrsize=\epsfury\pspoints
   \advance\epsfrsize by-\epsflly\pspoints
   \epsftsize=\epsfurx\pspoints
   \advance\epsftsize by-\epsfllx\pspoints
   \epsfxsize=\epsfsize{\epsftsize}{\epsfrsize}%
   \ifnum \epsfxsize=0
      \ifnum \epsfysize=0
        \epsfxsize=\epsftsize
        \epsfysize=\epsfrsize
        \epsfrsize=0pt
      \else
        \epsftmp=\epsftsize \divide\epsftmp\epsfrsize
        \epsfxsize=\epsfysize \multiply\epsfxsize\epsftmp
        \multiply\epsftmp\epsfrsize \advance\epsftsize-\epsftmp
        \epsftmp=\epsfysize
        \loop \advance\epsftsize\epsftsize \divide\epsftmp 2
        \ifnum \epsftmp>0
           \ifnum \epsftsize<\epsfrsize
           \else
              \advance\epsftsize-\epsfrsize \advance\epsfxsize\epsftmp
           \fi
        \repeat
        \epsfrsize=0pt
      \fi
   \else
     \ifnum \epsfysize=0
       \epsftmp=\epsfrsize \divide\epsftmp\epsftsize
       \epsfysize=\epsfxsize \multiply\epsfysize\epsftmp
       \multiply\epsftmp\epsftsize \advance\epsfrsize-\epsftmp
       \epsftmp=\epsfxsize
       \loop \advance\epsfrsize\epsfrsize \divide\epsftmp 2
       \ifnum \epsftmp>0
          \ifnum \epsfrsize<\epsftsize
          \else
             \advance\epsfrsize-\epsftsize \advance\epsfysize\epsftmp
          \fi
       \repeat
       \epsfrsize=0pt
     \else
       \epsfrsize=\epsfysize
     \fi
   \fi
}%
\def\epsfstatus#1{
   \ifepsfverbose
     \immediate\write16{#1: BoundingBox:
                  llx = \epsfllx\space lly = \epsflly\space
                  urx = \epsfurx\space ury = \epsfury\space}%
     \immediate\write16{#1: scaled width = \the\epsfxsize\space
                  scaled height = \the\epsfysize}%
   \fi
}%
{\catcode`\%=12 \global\let\epsfpercent=
\global\def\epsfatend{(atend)}%
\long\def\epsfaux#1#2:#3\\%
{%
   \def\testit{#2}
   \ifx#1\epsfpercent           
       \ifx\testit\epsfbblit    
            \epsfgrab #3 . . . \\%
            \ifx\epsfllx\epsfatend 
                \global\epsfatendtrue
            \else               
                \ifepsfatend    
                \else           
                    \epsffileokfalse
                \fi
                \global\epsfbbfoundtrue
            \fi
       \fi
   \fi
}%
\def\epsfempty{}%
\def\epsfgrab #1 #2 #3 #4 #5\\{%
   \global\def\epsfllx{#1}\ifx\epsfllx\epsfempty
      \epsfgrab #2 #3 #4 #5 .\\\else
   \global\def\epsflly{#2}%
   \global\def\epsfurx{#3}\global\def\epsfury{#4}\fi
}%
\def\epsfsize#1#2{\epsfxsize}%